# Cryptographic Hardening of d-Sequences

Sandhya Rangineni

**Abstract:** This paper shows how a one-way mapping using majority information on adjacent bits will improve the randomness of d-sequences.

## Introduction

In a recent article, Anthes [1] summarizes recent results in generating true random numbers. In particular, it mentions the work at Intel Corp. to use thermal noise on the central processing unit of the computer as random number generator (RNG) [2]. This is not unlike the RNGs based on quantum processes that have been proposed elsewhere [3]. Quantum processes come with their own uncertainty [4]-[7].

Randomness is generally measured in terms of probability or of complexity. From the lens of probability, all binary sequences of length *n* are equivalent. From the point of view of complexity, randomness will depend on the algorithm that has been used to generate the sequence. Ritter provides a summary of several measures of algorithmic complexity [8] and, therefore, also of randomness.

In this article, we will investigate results of a method of cryptographic strengthening of RNGs. Basically, the idea is to apply a many–to-one mapping to the binary output of the RNG, increasing the complexity of reverse process. We show that by using a 3-to-1 mapping where each group of three 0s and 1s is replaced by whatever the majority improves the autocorrelation function of the resultant sequence in some cases. This will be tried both for the Windows based RNGs as well as d-sequences [9-17], that are "decimal sequences" in an arbitrary base, although binary (base-2) sequences are the ones considered here. D-sequences have found several applications in cryptography and they are of particular interest since any random sequence can be represented as a d-sequence (Figure 1).

## Randomness measured by Autocorrelation Function

For simplicity, we consider only the autocorrelation function as measure of randomness. The value of the autocorrelation is defined as in the equation below:

$$C(k) = \frac{1}{n} \sum_{j=0}^{n} a_j a_{j+k}$$

A good random sequence has an autocorrelation function that is roughly two-valued. The *C(k)* function for a maximum-length d-sequence is has a negative peak of -1 for half the period because of the anti-symmetry of the sequence. Non-maximum length d-sequences need not have such a structure.

The binary d-sequence is generated by means of the algorithm [11]:

$$a(0) = 2$$
$$b(i+1) = 2b(i) \bmod q$$
$$a(i) = b(i) \bmod 2$$

where *q* is a prime number. The maximum length (with period q-1) sequences are generated when 2 is a primitive root of *q*. When the binary d-sequence is of maximum length, then bits in the second half of the period are the complements of those in the first half.

Any periodic sequence can be represented as a generalized d-sequence *m/n*, where *m* and *n* are suitable natural numbers, i.e., positive integers.

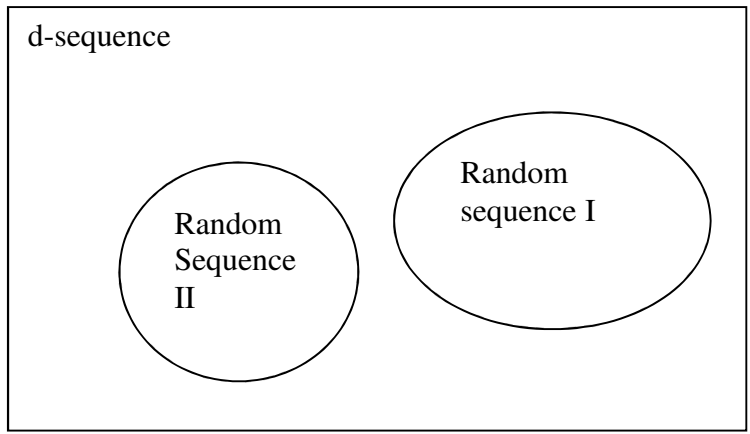

**Figure 1:** General random sequences as a subset of d-sequences

## The PR(n) sequence

The PR(n) sequences emerges by mapping each group of adjacent n bits (n odd) of the PR sequence to 0 or 1 depending on whether it has a majority of 0s or 1s. We have done experiments on many d-sequences (see below for examples) and we find that PR(3) provides significant improvement and that there is no significant advantage in taking larger values of n.

The tables below provide the list of a few of the largest values of the autocorrelation function for the given sequence. The off -1 or -1 values are 0.33 and -0.33. We see that these values have reduced to 0.11 and -0.13 for the d-sequence corresponding to 1907. There is corresponding improvement (not necessarily of the same extent) for the other examples given below.

**Table 1.** Reduction in the largest value autocorrelation values for a few d-sequences

### 1907:

| PR(1) | PR(3) | PR(5) | PR(7) | PR(9) | PR(11) |
|---|---|---|---|---|---|
| 1.0 | 1.0 | 1.0 | 1.0 | 1.0 | 1.0 |
| -1.0 | -0.50 | -0.37 | -0.46 | -0.81 | -0.38 |
| 0.33 | 0.11 | 0.11 | 0.18 | 0.19 | 0.26 |
| -0.33 | -0.13 | -0.12 | -0.13 | -0.21 | -0.17 |
| 0.20 | 0.08 | 0.10 | 0.12 | 0.18 | 0.18 |
| -0.20 | -0.10 | -0.11 | -0.12 | -0.20 | -0.16 |
| 0.14 | 0.07 | 0.09 | 0.10 | 0.17 | 0.17 |
| -0.14 | -0.07 | -0.10 | -0.10 | -0.18 | -0.13 |

### 2243:

| PR(1) | PR(3) | PR(5) | PR(7) | PR(9) | PR(11) |
|---|---|---|---|---|---|
| 1.0 | 1.0 | 1.0 | 1.0 | 1.0 | 1.0 |
| -1.0 | -0.49 | -0.38 | -0.42 | -0.39 | -0.80 |
| 0.33 | 0.19 | 0.12 | 0.11 | 0.12 | 0.2 |
| -0.33 | -0.12 | -0.09 | -0.19 | -0.21 | -0.2 |
| 0.19 | 0.09 | 0.11 | 0.10 | 0.11 | 0.19 |
| -0.19 | -0.08 | -0.07 | -0.11 | -0.20 | -0.19 |
| 0.14 | 0.08 | 0.10 | 0.09 | 0.10 | 0.14 |
| -0.14 | -0.07 | -0.06 | -0.09 | -0.19 | -0.14 |

### 2333:

| PR(1) | PR(3) | PR(5) | PR(7) | PR(9) | PR(11) |
|---|---|---|---|---|---|
| 1.0 | 1.0 | 1.0 | 1.0 | 1.0 | 1.0 |
| -1.0 | -0.52 | -0.40 | -0.39 | -0.35 | -1.0 |
| 0.33 | 0.09 | 1.13 | 0.12 | 0.15 | 0.21 |
| -0.33 | -0.10 | -0.11 | -0.17 | -0.21 | -0.21 |
| 0.19 | 0.06 | 0.08 | 0.11 | 0.14 | 0.19 |
| -0.19 | -0.09 | -0.09 | -0.14 | -0.18 | -0.19 |
| 0.14 | 0.05 | 0.07 | 0.10 | 0.12 | 0.18 |
| -0.14 | -0.08 | -0.08 | -0.13 | -0.15 | -0.18 |

It is interesting that the performance of PR(n) for a larger value of n does not necessarily imply improved results as far as the autocorrelation function is concerned. For the sake of illustration, we now present the autocorrelation functions for PR(1), PR(3), PR(7), and PR(11) obtained using the d-sequence: 1/1571 .

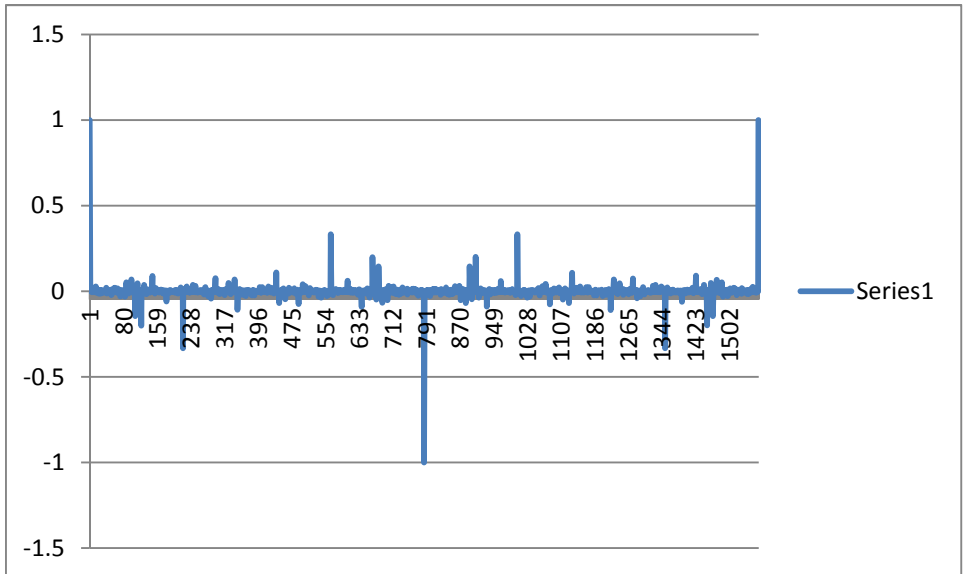
**Figure 2**. Autocorrelation function for PR(1) for 1/1571

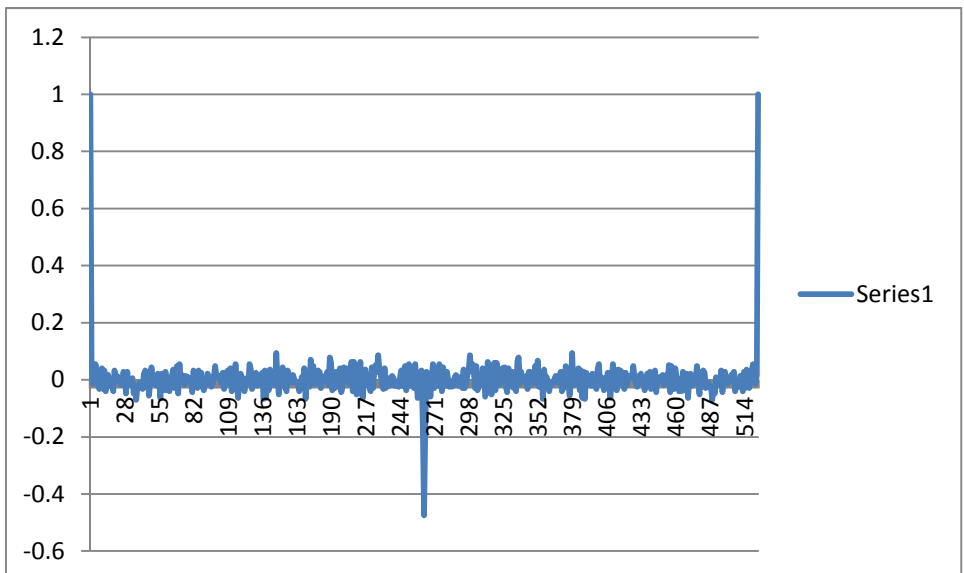
**Figure 3**. Autocorrelation function for PR(3) for 1/1571.

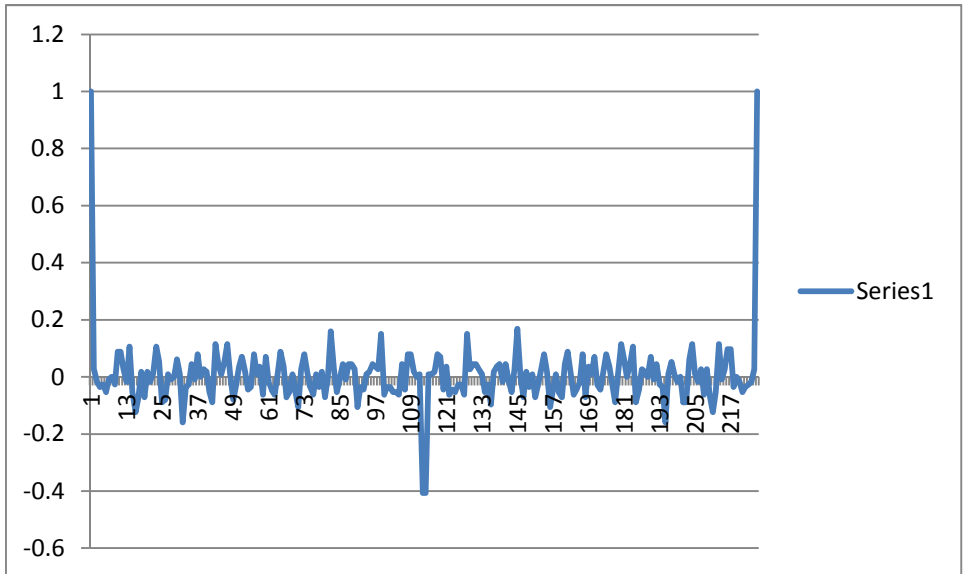
**Figure 4**. Autocorrelation function for PR(7) for 1/1571.

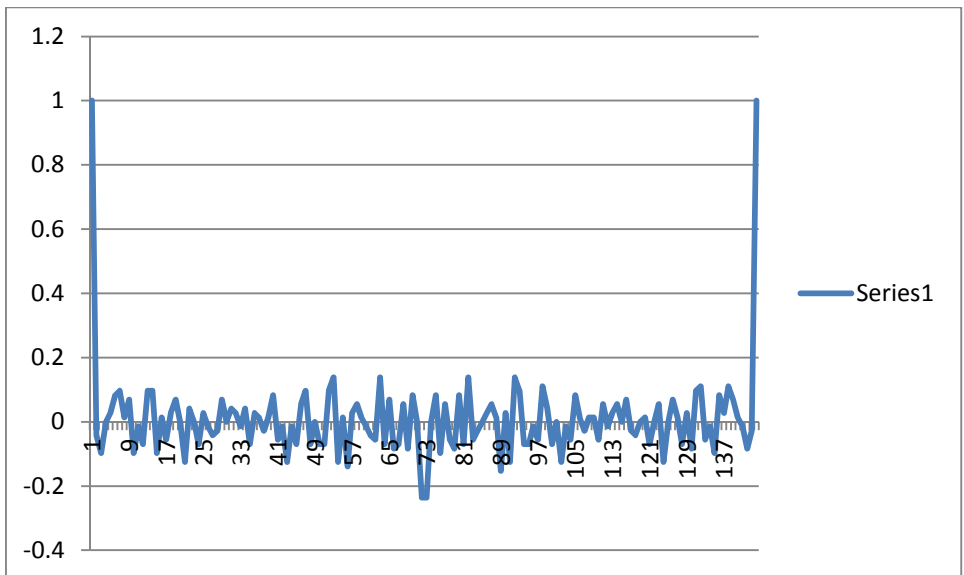
**Figure 5**. Autocorrelation function for PR(11) for 1/1571.

Note that the negative peak for half the period gets smaller and smaller as we increase n in PR(n). This shows that the improvement in randomness can be quite dramatic.

## Random Numbers from Windows PC

In a similar way shown above, I have taken binary random sequences generated by using random number generators in windows PC and done some experiments in order to determine how many-to-one mapping impacts the quality of the autocorrelation function.

The results for the Windows PC RNG are shown in Table 2.

Table 2. The largest value autocorrelation
values for Windows RNG

| PR(1) | PR(3) | PR(5) | PR(7) | PR(9) | PR(11) |
|---|---|---|---|---|---|
| 1.0 | 1.0 | 1.0 | 1.0 | 1.0 | 1.0 |
| -0.07 | -0.12 | -0.15 | -0.19 | -0.20 | -0.16 |
| 0.06 | 0.10 | 0.14 | 0.18 | 0.16 | 0.19 |
| -0.06 | -0.10 | -0.14 | -0.15 | -0.14 | -0.15 |
| 0.05 | 0.09 | 0.12 | 0.16 | 0.14 | 0.18 |
| -0.05 | -0.09 | -0.13 | -0.14 | -0.13 | -0.12 |
| 0.04 | 0.08 | 0.11 | 0.15 | 0.13 | 0.13 |
| -0.04 | -0.08 | -0.10 | -0.13 | -0.12 | -0.11 |

Note that in this case the use of applying the many-to-one mapping does not improve the autocorrelation function of this RNG.

## PR(n) mapping applied to the Mesh random sequence

We now apply the PR(n) operation to the ransom sequence from the mesh array [20; also see [15]-[19]). The results are as follows:

Table 3. The largest value autocorrelation
values for the mesh array sequence

| PR(1) | PR(3) | PR(5) | PR(7) | PR(9) | PR(11) |
|---|---|---|---|---|---|
| 1.0 | 1.0 | 1.0 | 1.0 | 1.0 | 1.0 |
| -0.08 | -0.14 | -0.22 | -0.18 | -0.19 | -0.23 |
| 0.08 | 0.12 | 0.22 | 0.19 | 0.19 | 0.17 |
| -0.07 | -0.13 | -0.16 | -0.16 | -0.17 | -0.15 |
| 0.07 | 0.11 | 0.18 | 0.18 | 0.18 | 0.15 |
| -0.06 | -0.11 | -0.15 | -0.15 | -0.16 | -0.13 |
| 0.06 | 0.10 | 0.14 | 0.15 | 0.15 | 0.13 |
| -0.05 | -0.10 | -0.12 | -0.13 | -0.14 | -0.10 |

The performance of the mesh array sequence to the many-to-one mapping is similar to that for the Windows RNG.

## Nested PR(n) sequences

It is significant that the performance of nested PR(n) sequences for the Windows PRNG is not very good as given by the results in Table 4 below.

**Table 4.** Nested values for Windows RNG

| PR(1) | PR(3) | PR3(3) | PR3(PR3(3)) |
|---|---|---|---|
| 1.0 | 1.0 | 1.0 | 1.0 |
| -0.07 | -0.12 | -0.12 | -0.21 |
| 0.06 | 0.10 | 0.15 | 0.18 |
| -0.06 | -0.10 | -0.10 | -0.18 |
| 0.05 | 0.09 | 0.14 | 0.16 |
| -0.05 | -0.09 | -0.09 | -0.13 |
| 0.04 | 0.08 | 0.13 | 0.13 |
| -0.04 | -0.08 | -0.08 | -0.10 |

## Conclusions

This article shows that the many-to-one mapping improves the quality of the autocorrelation function of d-sequences. Since this mapping does not correspondingly improve the performance of the Windows RNG or the recently introduced mesh array sequence, this mapping could have applications in the evaluation of the quality of randomness of a sequence.

## References

1. G. Anthes, The quest for randomness. Communications of the ACM, 54: 13-15, 2011.

2. S. Srinivasan et al. 2010 IEEE Symposium on VLSI Circuits, June 16-18, 2010.

3. M.A.Nielsen and I.L. Chuang, Quantum Computation and Quantum Information. Cambridge University Press, 2000.

4. R. Landauer, The physical nature of information. Phys. Lett. A 217, 188-193, 1996.

5. S. Kak, Information, physics and computation. Foundations of Physics 26, 127-137, 1996.

6. S. Kak, The initialization problem in quantum computing. Foundations of Physics 29, pp. 267-279, 1999.

7. S. Kak, Quantum information and entropy. International Journal of Theoretical Physics 46, 860-876, 2007.

8. T. Ritter, Randomness tests: a literature survey. http://www.ciphersbyritter.com/RES/RANDTEST.HTM

9. S. Kak and A. Chatterjee, On decimal sequences. IEEE Transactions on Information Theory, IT-27: 647 – 652, 1981.

10. S. Kak, Encryption and error-correction coding using D sequences. IEEE Transactions on Computers, C-34: 803-809, 1985.

11. S. Kak, New results on d-sequences. Electronics Letters, 23: 617, 1987.

12. D. Mandelbaum, On subsequences of arithmetic sequences. IEEE Trans on Computers, vol. 37, pp 1314-1315, 1988.

13. N. Mandhani and S. Kak, Watermarking using decimal sequences. Cryptologia, vol 29, pp. 50-58, 2005; arXiv: cs.CR/0602003

14. K. Penumarthi and S. Kak, Augmented watermarking. Cryptologia, vol. 30, pp 173-180, 2006.

15. S. Kak, Multilayered array computing. Information Sciences 45, 347-365, 1988.

16. S. Kak, A two-layered mesh array for matrix multiplication. Parallel Computing 6, 383-385, 1988.

17. S. Kak, On the mesh array for matrix multiplication. 2010. arXiv:1010.5421

18. S. Kak, An overview of analog encryption. Proceedings IEE 130, Pt. F, 399-404, 1983.

19. S. Kak and N.S. Jayant, On speech encryption using waveform scrambling. Bell System Technical Journal 56, 781-808, 1977.

20. S. Rangineni, New results on scrambling using the mesh array. arXiv:1102.4579